\documentclass[prl,aps,amsmath,amssymb,lengthcheck,superscriptaddress,floatfix]{revtex4}
\usepackage{graphicx}
\usepackage{dcolumn}
\usepackage{bm}
\usepackage[usenames,dvipsnames]{color}
\usepackage{psfrag}

\begin{document}

\title{Dynamical bifurcation as a semiclassical counterpart of a quantum phase transition}

\author{P. Buonsante}
\affiliation{Dipartimento di Fisica, Universit\`a degli Studi di Parma, V.le G.P. Usberti n.7/A, 43100 Parma, Italy}
\author{V. Penna}
 \affiliation{Dipartimento di Fisica, Politecnico di Torino, Corso Duca degli Abruzzi 24, I-10129 Torino, Italy}%
 \affiliation{CNISM, u.d.r. Politecnico di Torino, Corso Duca degli Abruzzi 24, I-10129 Torino, Italy}%
\author{A. Vezzani}
 \affiliation{Centro S3, CNR Istituto di Nanoscienze	, via Campi 213/a, 41100 Modena, Italy}
\affiliation{Dipartimento di Fisica, Universit\`a degli Studi di Parma, V.le G.P. Usberti n.7/A, 43100 Parma, Italy}
\date{\today}

\begin{abstract}
We illustrate how  dynamical transitions in nonlinear semiclassical models can be recognized as phase transitions in the corresponding -- inherently linear -- quantum model, where,  in a Statistical Mechanics framework, the thermodynamic limit is realized  by letting the particle population go to infinity at fixed size. We focus on lattice bosons described by the  Bose-Hubbard (BH) model and Discrete Self-Trapping (DST) equations at the quantum and semiclassical level, respectively.
 After showing  that  the gaussianity of the quantum ground states is broken at the phase transition, 
we evaluate finite populations effects introducing a suitable scaling hypothesis; we work out the exact value of the critical exponents and we provide numerical evidences confirming our hypothesis. 
Our analytical results rely on a general scheme obtained from a large-population expansion of the eigenvalue equation of the BH model. In this approach the DST equations resurface as solutions of the zeroth-order problem.
\end{abstract}
\maketitle

The emergence of nonlinear phenomena typical of semiclassical models  from more fundamental, inherently linear quantum models represents a basic question which has attracted a fair amount of attention in the literature.
The large number of experiments with ultracold atoms trapped in optical lattices evidencing both quantum \cite{Greiner_Nature_415_39,Sherson_Nature_467_68,Paredes_Nature_429_277,Fallani_PRL_98_130404} and nonlinear \cite{Cataliotti_Science_293_843,Anker_PRL_94_020403,Gati_PRL_95_010402,Winkler_Nature_441_853}  effects make the  the Bose-Hubbard model \cite{Fisher_PRB_40_546,Jaksch_PRL_81_3108,Bloch_NatPh_1_23} an important playground in this respect.
A general form for the Bose-Hubbard Hamiltonian, describing interacting bosons hopping on the sites of a discrete structure, is
\begin{equation}
\label{BH}
H = \sum_{j=1}^L \left[ \frac{sU}{2} \big(a_j^\dag\big)^2\big(a_j\big)^2+v_j n_j \right]\! -\! J \sum_{j \ell} a_j^\dag A_{j \ell} a_\ell
\end{equation}
where $j$ is a site label, $A_{j\,\ell}$ is the adjacency matrix of the structure,  $a_j$ is an on-site (annihilation) boson operator, $[a_j,\, a_\ell^\dag]=\delta_{j\ell}$. As to the parameters, $U>0$ is the strength of the on-site interaction, whose attractive or repulsive character is dictated by $s=\pm 1$, and $J$ is the hopping amplitude. The semiclassical counterpart of Eq.~\eqref{BH} 
\begin{equation}
\label{DSTH}
{\cal H} = \sum_{j=1}^L \left[ \frac{s U}{2} |\alpha_j|^4 +v_j |\alpha_j|^2 \right]\! -\! J \sum_{j \ell} \alpha_j^* A_{j \ell} \alpha_\ell
\end{equation}
can be obtained by trading the quantum operator $a_j$ for a C-number $\alpha_j$, whose  square modulus and angle represent the boson population and macroscopic phase attached to  site $j$, respectively. 
Actually,  the complex variables $\alpha_j$ governed by Hamiltonian \eqref{DSTH} are the variational parameters of a coherent-state ansatz for Hamiltonian \eqref{BH} \cite{Wright_PhisicaD_69_18,Amico_PRL_80_2189,Buonsante_PRA_72_043620}. 
The nonlinear equation of motion ensuing from Hamiltonian \eqref{DSTH} are known as {\it discrete self-trapping} (DST) equations \cite{Eilbeck_PhysicaD_16_318}, and the relevant normal modes are often compared to the eigenstates of Hamiltonian \eqref{BH}.

One of the most striking features stemming from  the nonlinear character of DST equations  is the occurrence of dynamical instabilities, such as modulational instability \cite{Eilbeck_PhysicaD_16_318}:
 for (attractive) interactions among bosons exceeding a critical value, the uniform solution of the DST equations on translation invariant lattices becomes unstable. This critical value is expected to coincide with the threshold for spatial localization in the ground-state of the system (i.e. soliton formation).
The quantum counterpart of this well known semiclassical feature has been investigated --- mostly on two- and three-site lattices at finite population --- by employing several indicators, such as energy gaps \cite{Cirac_PRA_57_1208}, number fluctuations \cite{Spekkens_PRA_59_3868,Ho_JLTP_135_257,Jack_PRA_71_023610,Zin2}  condensate fraction \cite{Buonsante_PRA_72_043620}, occupation probability in the Fock space \cite{Javanainen_PRL_101_170405},  {\it localization width} \cite{Buonsante_PRA_82_043615}, fidelity \cite{Oelkers_PRB_75_115119},  entanglement \cite{Pan_PLA_339_403,Fu_PRA_74_063614,Viscondi2,Note1}, Bethe ansatz techniques \cite{Pan_PLA_339_403,Dunning_JSM_P11005}.
Here we show that  this semiclassical dynamical transition can be seen as a genuine quantum phase transition -- characterized by a vanishing gap -- in which the {\it thermodynamic limit} is 
realized at fixed lattice size by letting the  bosonic population go to infinity. We connect the order of the phase transitions to the bifurcation pattern characterizing the solutions to the DST equations.
Specifically,  we find that on 1D lattices comprising $L=2$ and $L\geq 6$ sites the modulational instability threshold coincides with a second-order critical point, while  for $L=3,4,5$ only  first-order transitions are present. 
In the second-order case, we determine the critical exponents characterizing the divergence of the fluctuations. 
Also, we analyze the crossover heralding the phase transition at finite population by verifying the finite-size  scaling hypothesis typical of Statistical Mechanics. In the same framework we show that at finite population the spectrum is not gapless at the transition point, but exhibits the expected avoided level crossing pattern.

In carrying out our analysis we develop a fully analytical general scheme based on a large-population expansion of the secular equation for the BH Hamiltonian on a $L$-site lattice. This expansion results in  a Schr\"odinger-like equation (SLe) set in a $L$-dimensional space directly related to the Fock space of the original quantum problem. In this framework the DST equations emerge naturally  as the equations for the extrema of  the potential part of the SLe, at the lowest order of the expansion.
The analysis of the {\it small oscillations} about the local minima of such potential gives access to an entire class of solutions of the original quantum problem,
which improve significantly on the coherent states employed as  trial wave functions in the variational derivation of the DST equations.
Similar to the coherent state, away from criticality our solution is a Gaussian function centered at a particular occupation-number Fock state dictated by the time-independent DST equations.  However, our perturbative approach provides an exact calculation of  the Gaussians widths. This gives us access to the correct quantum fluctuations in the system.
Furthermore, our picture captures situations where the solutions of the SLe are not Gaussian, which correspond to the above mentioned quantum critical points.

Our analysis starts from the eigenvalue equation for the BH model, $H |\Psi\rangle=E |\Psi\rangle$, where the eigenstate $|\Psi\rangle$ has been expanded over
the Fock space of occupation numbers,
\begin{equation}
\label{FockExp}
|\Psi\rangle = {\sum_{\vec {x}}}' \psi(\vec x) |\vec x\rangle, \quad |\vec x\rangle = \prod_{j=1}^L  \frac{\left(a_j^\dag\right)^{N x_j}}{(N x_j)!}|0\rangle.
\end{equation}
The labels of the Fock state $|\vec{x}\rangle$ have been conveniently normalized, so that $x_j \in [0, 1]$ independent of the total boson population $N$, which is conserved owing to the  commutation relation $[H,\sum_j a_j^\dag a_j]=0$. The prime on the summation symbol in the first of Eqs.~\eqref{FockExp} signals that $x_j = k_j N^{-1}$, with $k_j\in {\mathbb N}_0$, and $\sum_j x_j =1$. When recast in terms of the expansion coefficients $\psi(\vec{x})$, the eigenvalue equation can be seen as a discrete equation on the vertices of a mesh grid covering the portion of $(L-1)$--dimensional hyperplane defined by $\sum_j x_j =1$ and $x_j \in [0,\, 1]$. The $N\to \infty$ limit plays the twofold role of a {\it thermodynamic limit} for the size of the fixed-number Fock space, and of a continuous limit for the mesh grid. Assuming that the expansion coefficients can be seen as a continuous function, $\psi(x_1,\,x_2,\cdots,x_\ell+N^{-1},\cdots,x_N)-\psi(x_1,\,x_2,\cdots,x_\ell,\cdots,x_N)=O(N^{-1})$ the eigenvalue equation becomes a SLe of the form $[{\cal U}-{\cal D}] \psi(\vec{x}) = \bar E \psi(\vec{x})$ where, to the leading order in $N^{-1}$, the self-adjoint operators are
\begin{align}
\label{qSEu}
{\cal U}(\vec{x}) & =  \sum_j \left(s x_j^2 +{\bar v}_j x_j\right) -2 \tau \sum_{j \ell} A_{j \ell}  \sqrt{x_j x_\ell} \\
\label{qSEd}
{\cal D}(\vec{x}) & = \frac{\tau}{N^2}  \sum_{j \ell} A_{j \ell} \frac{\partial}{\partial (x_j-x_\ell)} \sqrt{x_j x_\ell} \frac{\partial}{\partial (x_j-x_\ell)}  
\end{align}

The effective parameters appearing in Eqs.~\eqref{qSEu} and \eqref{qSEd} are $\tau = J/U N$, $\bar v_j = v_j/U N$ and $\bar E = E/U N^2$. We remark that a similar approach has been adopted in Refs.~\cite{Spekkens_PRA_59_3868,Javanainen_PRA_60_4902,Franzosi_PRA_63_043609,Ho_JLTP_135_257,Zin2}, mostly for the two-site case. Note that the leading terms  in ${\cal D}(\vec{x})$ and ${\cal U}(\vec{x})$ are  of order $N^0$ and $N^{-2}$, respectively. This suggests that significant solutions to the above differential equation might be strongly localized in the vicinity of the local minima of ${\cal U}(\vec{x})$. Thus, a satisfactory description of a subset of eigestates of Hamiltonian~\eqref{BH} can be obtained through the analysis of the ``small oscillations'' about such minima.

The stationarization of the leading term of  Eq.~\eqref{qSEu} results in the set of equations
\begin{equation}
\label{stat}
s x_j \sqrt{x_j}+ {\bar v}_j \sqrt{x_j} - \tau \sum_\ell A_{j \ell} \sqrt{x_\ell} = \lambda \sqrt{x_j}, 
\end{equation}
where $\lambda$ is a Lagrange multiplier enforcing the constraint on $\vec{x}$. Note that, upon setting $\sqrt{x_j} = \alpha_j$, Eqs.~\eqref{stat} are  equivalent to the time-independent DST equations, which thus emerge in our description without any variational ansatz.

\begin{figure}[t]
\includegraphics[height=3.7cm]{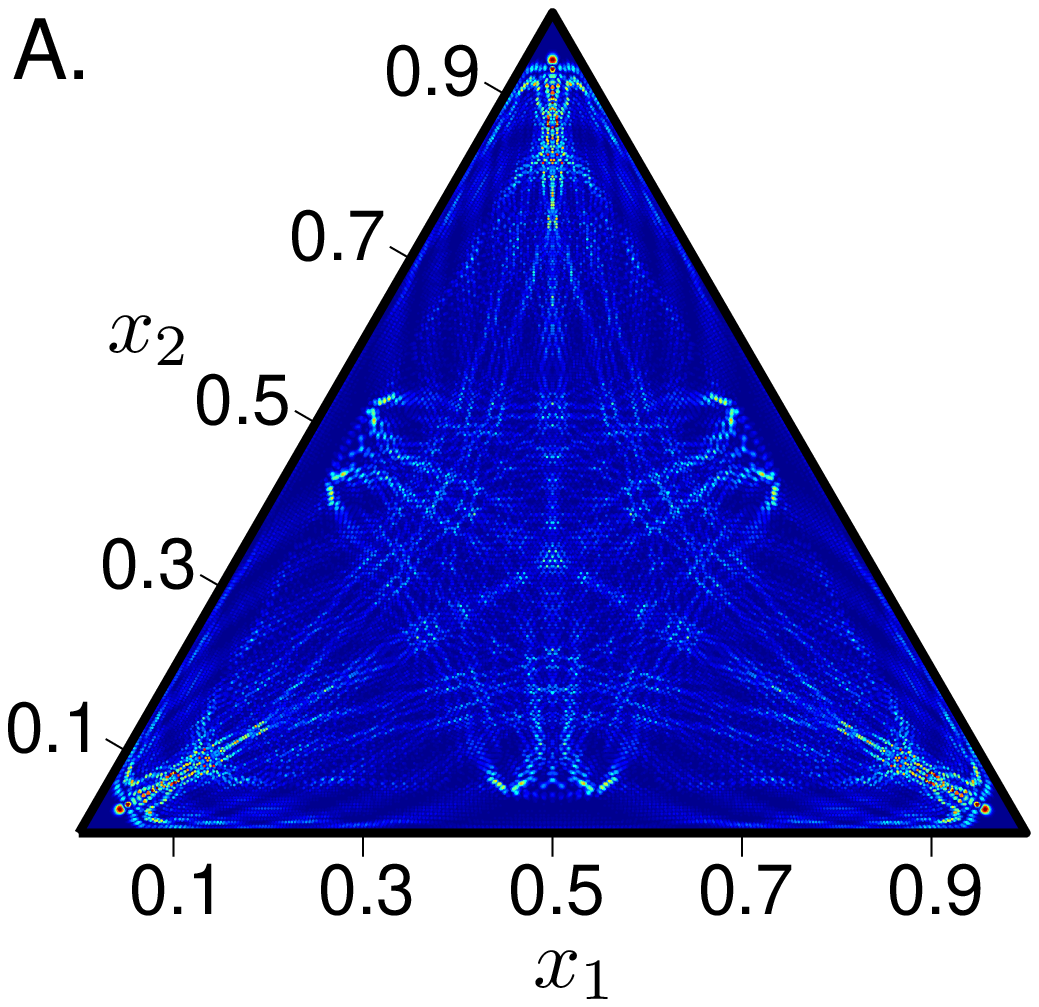}
\includegraphics[height=3.7cm,bb = 138 300 382 545,clip]{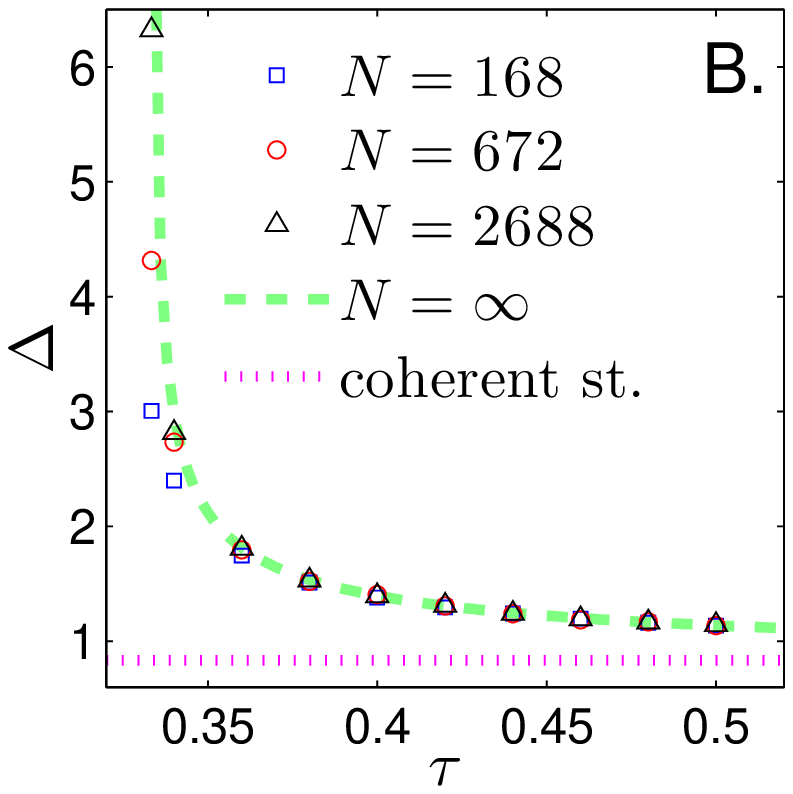}
\caption{\label{trimer0} (color online) A. density plot for $|\psi(x_1,x_2,1-x_1-x_2)|^2$ for a mid-spectrum eigenstate of a $L=3$ lattice containing $N=600$ bosons, at $\tau=0.26>\tau_{\rm loc}$. The discontinuous character of the expansion coefficients is apparent; B. Normalized number fluctuations vs. effective hopping amplitude on a lattice comprising $L=6$ sites for increasing $N$. The thick dashed and dotted lines are the theoretical prediction obtained from Eq.~\eqref{sig} and from the coherent state approach \cite{Wright_PhisicaD_69_18,Amico_PRL_80_2189,Buonsante_PRA_72_043620}.}
\end{figure}

If the above stationary points  correspond to non-degenerate locally quadratic minima of ${\cal U}$, the problem is mapped onto a quantum harmonic oscillator. In particular, the expansion coefficients $\psi(\vec{x})$ of the (local) ground state is a Gaussian whose width is in principle accessible. 
 As we mention the trial-wavefunction in the coherent-state approach of Refs. \cite{Amico_PRL_80_2189,Buonsante_PRA_82_043615} is also a Gaussian, whose center  is  the Fock state corresponding to the relevant stationary  solution of the DST equations. However, once the center is chosen, the width of such Gaussian is constrained by the coherent-state structure.
Conversely, in our picture, the analysis of the local harmonic potential provides the correct width for the Gaussian function, which results the exact description of the scaling of quantum fluctuations in the $N\to \infty$ limit.
Also, while coherent trial wavefunctions are inherently Gaussian, our description allows for non-Gaussian eigenstates. It should be remarked that while our approach is surely applicable to low-lying eigenfunction of Hamiltonian \eqref{BH},  the assumption of continuity may not hold for higher energies, as exemplified in Fig.~\ref{trimer0}.


We now specialize the above discussion to the study of the localization/delocalization transition for attractive boson  on  1D, translation-invariant ($v_j=0$) lattice comprising $L$ sites. As it is well known, the uniform state $x_j = L^{-1}$ is always a solution to Eqs.~\eqref{stat}, and  it coincides with the absolute minimum of  ${\cal U}(\vec{x})$ \cite{Buonsante_PRE_75_016212,Buonsante_PRA_82_043615} for attractive interactions and $\tau>\tau_{\rm loc}$ as well as for repulsive interactions at any $\tau$. Below such {\it localization} threshold the minimum of ${\cal U}(\vec{x})$ is $L-$fold degenerate and the corresponding solutions of Eqs.~\eqref{stat} spontaneously break the translational symmetry due to the nonlinear interaction term. For $L>5$ the localization threshold coincides with the critical value below which the uniform state becomes modulationally unstable, $\tau_{\rm loc}=\tau_{\rm m.i.} = [2 L \sin^2(\pi/L)]^{-1}$. The same  happens in the two-site case. For $L=3,4,5$ the low lying stationary points of ${\cal U}(\vec{x})$ exhibit a more complex bifurcation pattern, and there is a region of metastability of the uniform state, $\tau_{\rm m.i.}<\tau_{\rm loc}$. As a consequence the $L$ equivalent, symmetry breaking local minima of ${\cal U}(\vec{x})$ do not merge continuously with the symmetric minimum at $\tau=\tau_{\rm loc}$, but disappear abruptly  \cite{Buonsante_PRE_75_016212,Buonsante_PRA_82_043615}. 
The analysis of the ground state of Eq.~\eqref{BH} is particularly simple in the region $\tau>\tau_{\rm loc}$, where the absolute minimum of ${\cal U}(x)$ is always locally harmonic. In order to analyze the small oscillations about such minimum  we introduce the small deviations $\eta_k$
\begin{equation}
\label{SmOs}
x_j = \frac{1}{L}+\frac{1}{\sqrt{L}} \sum_{k=1}^{L-1} \exp\left(i \frac{2\pi}{L} j k\right) \eta_k
\end{equation}
This choice decouples the leading terms of the 
equation $[{\cal U}-{\cal D}]\psi = \bar E \psi$ into $L-1$ independent equations, 
\begin{equation}
\label{HO}
\left[\frac{\eta_k^2}{4\sigma_k^4}-\frac{\partial^2}{\partial \eta_k^2}\right]\psi_k(\eta_k)=\bar{{\cal E}}_k \psi_k(\eta_k)
\end{equation}
where $\psi = \prod_k \psi_k$, $\bar E =4\tau/(L N)^2 \sum_k  {\cal E}_k/ \tau_k$, and
\begin{equation}
\label{sig}
\sigma_k^2(\tau) = \frac{1}{N L} \sqrt{\frac{\tau}{s \tau_{k} +\tau}}, \qquad \tau_k = \frac{1}{2 L \sin^2\left(\frac{\pi}{L} k\right)}
\end{equation}
In particular, the lowest energy gap of the original problem, Eq.~\eqref{BH}, is $\Delta E = 4J/L \sqrt{(s\tau_1+\tau)/(\tau \tau_1^2)}$.
We note that the above results were obtained in Ref.~\cite{Jack_PRA_71_023610} within a Bogoliubov approach and in Ref.~\cite{Javanainen_PRA_60_4902}, where Eq.~\eqref{HO} was derived introducing the small deviations from the uniform solution, Eq.~\eqref{SmOs}, directly into the structure of the Fock states.

Our method is however more  general, since it can be in principle applied to any stable solution of the time-independent DST equation, emerging as a non-degenerate locally harmonic minimum of ${\cal U}(\vec{x})$. Again, the eigenstate of Hamiltonian~\eqref{BH} is a Gaussian state whose square width vanishes as $N^{-1}$.
Note that, unlike Eq.~\eqref{sig}, the square width of the uniform solution  in the coherent-state approach \cite{Wright_PhisicaD_69_18,Amico_PRL_80_2189,Buonsante_PRA_72_043620} is $(NL)^{-1}$,  irrespective of $\tau$ and $k$.

 The above calculation allows us to evaluate,  the (normalized) number fluctuations $\Delta(\tau,N) = N (\langle x_j^2 \rangle-\langle x_j \rangle^2) = N \sum_{k\neq 0} \langle \eta_k^2 \rangle= N\sum_{k\neq 0} \sigma_k^2(\tau)$, where the factor $N$ ensures that this quantity remains finite in the large$-N$ limit. Panel B in Fig.~\ref{trimer0} shows that numerical estimates of $\Delta$   obtained from  {\it population} quantum Monte Carlo simulations \cite{Iba_TJSAI_16_279} exhibit a very satisfactory agreement with our analytical prediction in the large-$N$ limit.

\begin{figure}[t]
\includegraphics[height=3.7cm]{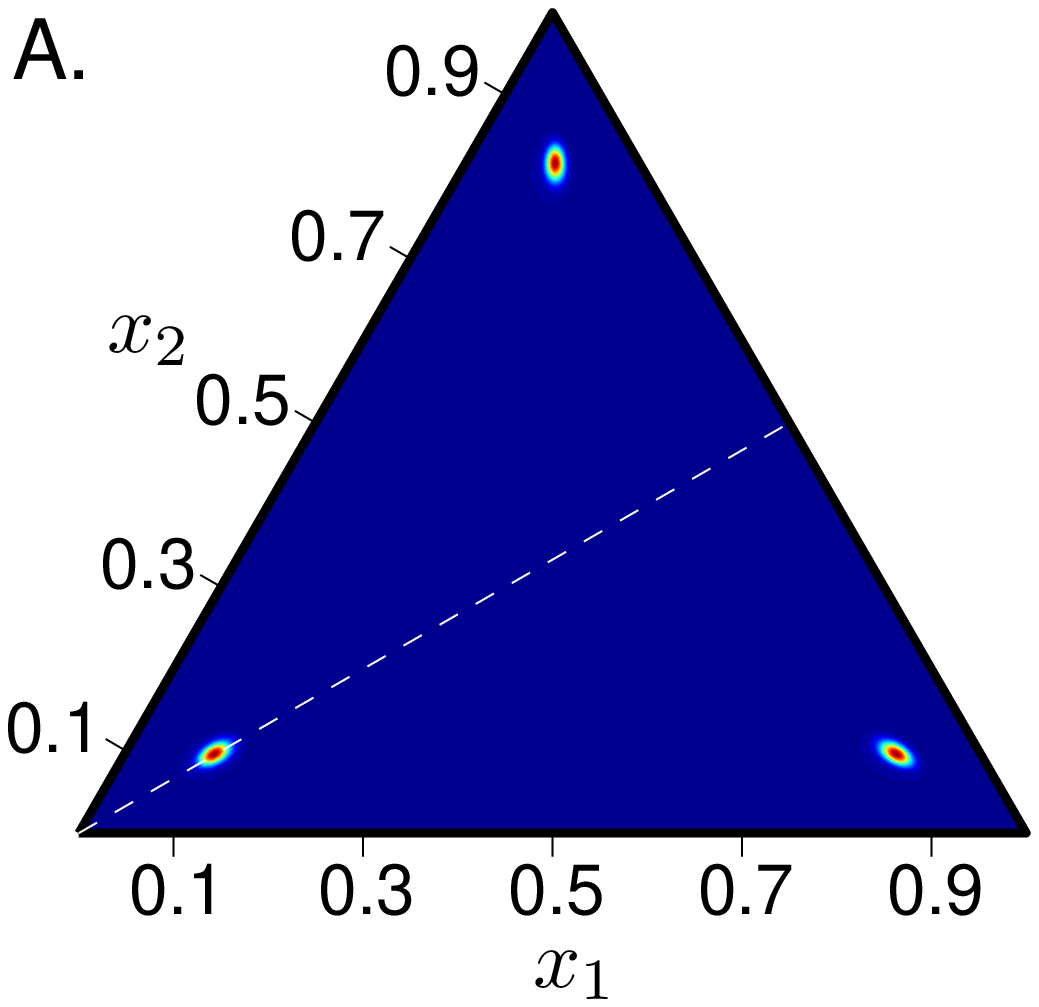}
\includegraphics[height=3.7cm,bb = 155 262 480 560,clip]{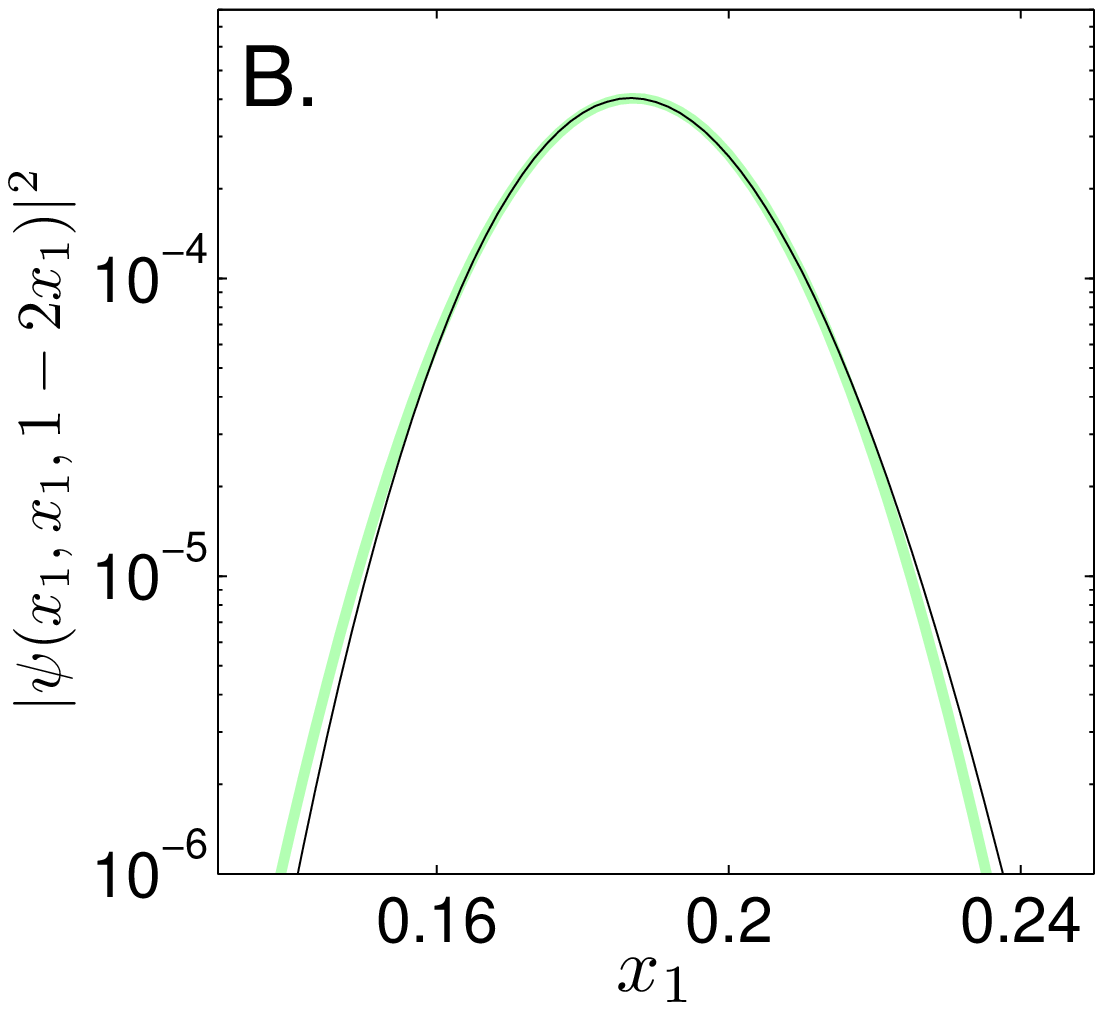}
\caption{\label{trimer1} (color online) Ground state for $N=1200$ attractive bosons on a lattice comprising $L=3$ sites, at $\tau = 0.21 < \tau_{\rm loc} = 0.25$. A. Density plot of $|\psi(x_1,x_2,1-x_1-x_2)|^2$  showing that that it consists of the superposition of three peaks. B. Plot along the cut $x_1=x_2$ (dashed white line in panel A.) showing that the peaks (black thin line) are well approximated by a Gaussian function (green thick line). The discrepancy at the tails is a finite-size (population) effect. See Ref.~\cite{Buonsante_PRA_82_043615} for more plots.}
\end{figure}

In the study of the ground state for attractive interactions ($s=-1$), two situations do not fit in the above general picture, and require a separate discussion. For $\tau<\tau_{\rm loc}$ the potential has $L$ degenerate (harmonic) absolute minima while for  $\tau = \tau_{\rm loc} = \tau_{\rm m.i.}$,  on lattices comprising $L=2$ or $L>5$ sites, the (non degenerate) absolute minimum of the potential is not harmonic. In the first case, the expansion coefficients $\psi(\vec x)$ are the superposition of $L$ Gaussian functions having the same width and centered at the absolute minimum points. Fig.~\ref{trimer1} illustrates an example of this situation for $L=3$. Thus the $\tau<\tau_{\rm loc}$ interval is analogous to a first-order phase transition line, where several different phases coexist. 

\begin{figure}[t!]
\includegraphics[width=\columnwidth,bb = 32 292 546 537,clip]{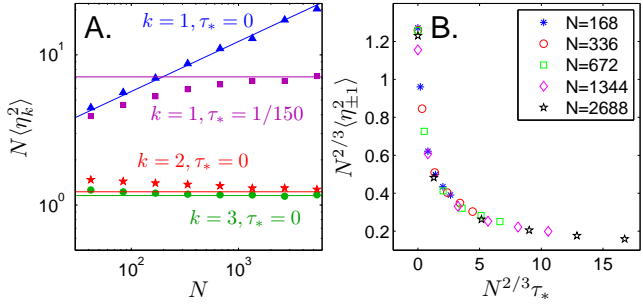}
\caption{\label{critp} (color online) A. Scaling of $N\langle \eta_k^2\rangle$ for $L=6$ and several $N$. The numeric data points (symbols) are compared to the theoretical predictions (solid lines) in Eqs.~\eqref{sig} and \eqref{critS} . B. Data collapse of finite$-N$ numerical data showing the correctness of the scaling hypothesis for the fluctuation of the first modes. }
\end{figure}

Likewise the $\tau = \tau_{\rm loc}=\tau_{\rm m.i.}$ case is analogous to a second-order critical point. Note indeed that, since $\tau_{\pm 1}=\tau_{\rm m.i.}$, the fluctuations of the first ``small oscillation'' modes diverge as $\sigma_{\pm 1}(\tau_*) \sim \tau_*^{-1/4}$  where $\tau_*=\tau-\tau_{\rm m.i.}$. Also the gap vanishes as $\Delta E \sim \tau_*^{1/2}$. This signals that the harmonic term of $\cal U$ vanishes for $\tau = \tau_{\rm m.i.}$ \cite{soft}, so that  the remaining leading term is quartic, and one has $[ \Gamma_{\pm 1} \eta_{\pm 1}^4 -N^{-2}\partial^2/\partial \eta_{\pm 1}^2] \psi_{\pm 1} = \bar { E}_{\pm 1} \psi_{\pm 1}$. Introducing the rescaled variables $\tilde \eta = N^{1/3} \eta_{\pm 1}$ and $\tilde {\cal E} = N^{4/3} \bar { E}_{\pm 1}$ we find that $\psi_{\pm 1}(\eta_{\pm 1}) = \varphi(N^{1/3}  \eta_{\pm 1})$, where $\varphi$ is  the (non-Gaussian) ground state solution of $[\Gamma_{\pm 1} \tilde \eta^4 -\partial^2/\partial \tilde \eta^2] \varphi = \tilde {\cal E} \varphi$. This means that, for $\tau_* = 0$ at finite $N$, the system is not gapless as suggested by the Gaussian approximation, but exhibits the expected avoided level crossing. Straightforward calculations allow us to determine the critical exponent governing the vanishing of the energy gap with increasing population, $\Delta E \propto J N^{-1/3}$. Likewise, we obtain   
\begin{equation}
\label{critS}
\langle\eta_{\pm 1}^2\rangle \sim N^{-2/3},  \qquad \langle\eta_{k}^2\rangle \sim N^{-1}\;\; {\rm for}\;\; |k|>1, 
\end{equation}
i.e. the singular part of the number fluctuation $\Delta$ is  dictated by $\langle \eta_{\pm 1}^2 \rangle$, where the scaling of the  $|k|>1$ modes is the same as far from criticality. Panel A. in Fig.~\ref{critp} illustrates the correctness of Eqs. \eqref{critS}  in the $L=6$ case.
Assuming that $\langle\eta_{\pm 1}^2\rangle$ obeys the standard finite-size scaling relation, $\langle\eta_{\pm 1}^2\rangle = N^{\gamma/\nu} \tilde f(N^{1/\nu}\tau_*)$, from Eqs.~\eqref{sig} and \eqref{critS} one easily works  out the critical exponents $\gamma = -1$ and $\nu = 3/2$. As we mention, here the boson population plays the role  reserved for the system size in the usual scaling approach.
Panel B. of Fig.~\ref{critp} illustrates how numerical data for a lattice comprising $L=6$ sites nicely collapse according to the above scaling hypothesis.
We conclude by remarking that for $2<L<6$, where $\tau_{\rm m.i.}<\tau_{\rm loc}$, there is no 2nd order critical point. The situation is similar to ferromagnetic Ising models with $p$-body interaction ($p>2$) \cite{Mouritsen_PRB_24_347}.

In summary, we develop a general technique for the study of bosonic quantum system in the large occupation limit. In this framework the corresponding semiclassical model is recovered without the need of an ansatz for the structure of the quantum state, and the dynamical transitions thereof incontrovertibly emerge as genuine quantum phase transitions. We study in detail the  localization transition of the paradigmatic Bose-Hubbard model with attractive interactions, for which we calculate explicitly some significant critical exponents and perform a finite-size scaling analysis. Our findings are supported by extensive numerical simulations.

\bibliography{./attractive.bib}

\begin{thebibliography}{34}
\bibitem[{sof()}]{soft}
\bibinfo{note}{That is, the lattice mode corresponding to $k=1$ in
  Eq.~\eqref{HO} goes soft.}
\end{thebibliography}

\end{document}